\documentclass[doublecol]{epl2} 

\title{Universal dynamic structure factor of a strongly correlated Fermi gas}

\author{H. Hu\inst{1}\thanks{hhu@swin.edu.au}\and X.-J. Liu\inst{1}\thanks{xiajiliu@swin.edu.au}}
\shortauthor{H. Hu and X.-J. Liu}

\institute{
\inst{1} ARC Centre of Excellence for Quantum-Atom Optics, \\
Centre for Atom Optics and Ultrafast Spectroscopy, \\
Swinburne University of Technology, Melbourne 3122, Australia}


\abstract{
Universality of strongly interacting fermions is a topic of great
interest in diverse fields. Here we investigate theoretically the
universal dynamic density response of resonantly interacting ultracold
Fermi atoms in the limit of either high temperature or large frequency: (1) 
At high temperature, we use quantum virial expansion to derive universal, 
non-perturbative expansion functions of dynamic structure factor;
(2) At large momentum, we identify that the second-order
expansion function gives the Wilson coefficient used in the operator
expansion product method. The dynamic structure
factor is therefore determined by its second-order expansion function
with an overall normalization factor given by Tan's contact parameter. 
We show that the spin parallel and antiparallel dynamic structure 
factors have respectively a tail of the form $\sim \pm \omega^{-5/2}$ for $\omega\rightarrow\infty$, 
decaying slower than the total dynamic structure factor found previously ($\sim \omega^{-7/2}$).
Our predictions for dynamic structure factor at high temperature or large frequency 
are testable using Bragg spectroscopy for ultracold atomic Fermi gases. }

\begin{document}

\maketitle

\section{Introduction}

The study of strongly interacting fermions has brought together very
different areas of physics - neutron stars, quark-gluon plasmas, high
temperature superconductors, and cold atoms - which, at first sight,
have little in common. There is, however, an important generic idea
of fermionic universality behind \cite{houniversality}: all strongly
interacting, dilute Fermi gases should behave identically, depending
only on scaling factors equal to the average particle separation and/or
thermal wavelength, but not on the details of the interaction. Recent
manipulation of ultracold Fermi gases of $^{6}$Li and $^{40}$K atoms
near a broad collisional (Feshbach) resonance provides an ideal avenue
to understand this fermionic universality \cite{giorgini,JILAEoS,dukeEoS,ensEoS}.
To date, universal thermodynamics of strongly interacting fermions
has been clearly demonstrated \cite{hdlNatPhys}, by measuring the
static equation of state. The purpose of this Letter is to show that
universality appears in \emph{dynamical} properties as well. We derive
\emph{exact} results for universal dynamic structure factor (DSF)
at high temperatures or at large momenta and frequencies.

Exact results are very valuable for strongly interacting fermions,
due to their non-perturbative, strongly correlated nature. There is
no small interaction parameter to control the accuracy of theories.
In specific cases, ab-initio calculations are possible using Monte
Carlo methods \cite{astrakharchik,bulgac,burovski}. However, in general
this approach suffers from the fermion sign problem \cite{akkineni}.
In 2008, Tan derived a set of exact, universal relations for two-component
(spin-1/2) Fermi gases with a large \textit{s}-wave scattering length
$a$ \cite{tan}. These universal relations all involve a many-body
parameter called the \emph{contact} ${\cal I}$, which measures the
density of pairs within short distances. Tan's relations can be understood
using the short-distance and/or short-time operator production expansion
(OPE) method \cite{braaten2008,bratten2010,son}, which separates
in a natural way few-body from many-body physics. For the {}``interaction''
DSF, $\Delta S_{\sigma\sigma^{\prime}}\left({\bf q},\omega,T\right)\equiv S_{\sigma\sigma^{\prime}}\left({\bf q},\omega,T\right)-S_{\sigma\sigma^{\prime}}^{(1)}({\bf q},\omega,T)$,
the OPE predicts that (${\bf q}\rightarrow\infty$ and $\omega\rightarrow\infty$)
\cite{bratten2010,son} 
\begin{equation}
\Delta S_{\sigma\sigma^{\prime}}\left({\bf q},\omega,T\right)\simeq W_{\sigma\sigma^{\prime}}\left({\bf q},\omega,T\right){\cal I}, \label{OPE}
\end{equation}
 where $(\sigma,\sigma^{\prime})=\uparrow,\downarrow$ denote the
spin, $S_{\sigma\sigma^{\prime}}$ and $S_{\sigma\sigma^{\prime}}^{(1)}$
are respectively the DSF of interacting and non-interacting Fermi
gases at the same chemical potential $\mu$ and temperature $T$,
and $W_{\sigma\sigma^{\prime}}$ are the temperature-dependent Wilson
coefficients that rely only on few-body physics. At high temperatures,
quantum virial expansion provides another rigorous means to bridge
few-body and many-body physics \cite{hovirial,horowitz,lhdprl2009,vecontact}.
It was shown that static thermodynamic properties of a strongly correlated
Fermi gas can be expanded non-perturbatively in fugacity using some
universal, temperature-independent virial coefficients \cite{ensEoS,lhdprl2009,vecontact},
which are exactly calculable from few-fermion solutions. Both OPE
and virial expansion give useful insight into the challenging many-body
problem. However, their connection is yet to be understood.

In this Letter, we investigate theoretically the universal dynamic properties of a strongly correlated
Fermi gas in the limit of either high temperature or large momentum/frequency. 
In the former limit, we show that the dynamic structure factor can be virial expanded 
in fugactiy, using some universal, temperature-implicit virial expansion functions. 
We derive, for the first time, these universal virial expansion functions for spin parallel and antiparallel DSFs
of a \emph{homogeneous} Fermi gas in the resonance (unitarity) limit,
where the scattering length diverges ($a\rightarrow\pm\infty$). In the latter limit of
large momentum, we show that the Wilson coefficient in Eq. (\ref{OPE}) is 
given by the second-order expansion function. Therefore, the large momentum
DSF is universally determined by the second order virial expansion, 
together with a many-body prefactor - the contact. Our results can be easily 
examined using Bragg spectroscopy for ultracold Fermi gases 
of $^6$Li or $^{40}$K atoms \cite{sutbragg}.

\section{Universal expansion function of DSF}

Virial expansion is a powerful tool for studying the high-temperature properties of ultracold
atomic Fermi gases \cite{lhdprl2009}. It expresses any physical quantities
of interest as an expansion in fugacity with some coefficients or
functions, since the fugacity $z\equiv\exp(\mu/k_{B}T)\ll1$ is a
controllable small parameter at high temperatures. For the interaction
DSF, the expansion is given by \cite{vedsf} \begin{equation}
\Delta S_{\sigma\sigma^{\prime}}\left({\bf q},\omega,T\right)=z^{2}\Delta S_{\sigma\sigma^{\prime},2}+z^{3}\Delta S_{\sigma\sigma^{\prime},3}+\cdots,\end{equation}
where $\Delta S_{\sigma\sigma^{\prime},n}({\bf q},\omega,T)$ ($n=2,3,\cdots$)
is the \textit{n}-th expansion function. The expansion is non-perturbative
as the few-body problem could be solved exactly, no matter how strong
the interaction. Virial expansion is anticipated to work for temperatures
down to the superfluid transition, although a nontrivial resummation
of expansion terms may be required if $z\gg1$. In Ref. \cite{vedsf},
the lowest second order expansion function $\Delta S_{\sigma\sigma^{\prime},2}^{(Trap)}$
for a \emph{trapped} Fermi gas was calculated using two-fermion solutions
in traps. However, the universal aspect of expansion functions was
not realized. As a result, for different temperatures/momenta the expansion
functions had to be re-calculated.

\begin{figure}
\onefigure[clip,width=0.48\textwidth]{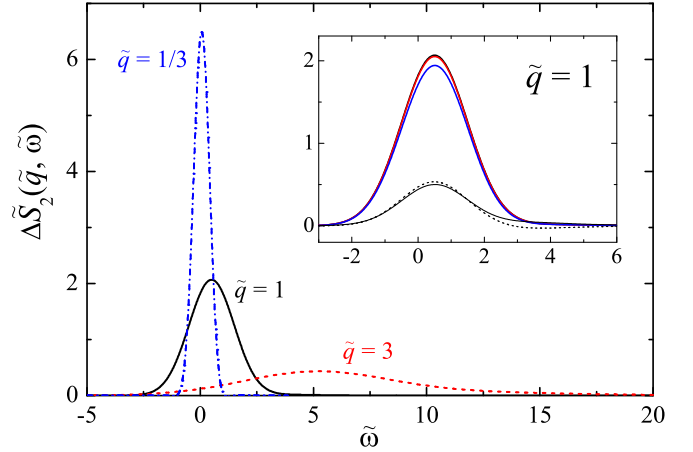} 
\caption{(Color on-line) Universal second order expansion function of DSF at
$\tilde{q}=1/3$, $1$, and $3$. The inset shows the rapid convergence
of $\Delta\tilde{S}_{2}(\tilde{q},\tilde{\omega})$ at small $\hbar\omega_{0}/(k_{B}T)$
(thick lines) and, $\Delta\tilde{S}_{\uparrow\uparrow,2}$ (thin solid
line) and $\Delta\tilde{S}_{\uparrow\downarrow,2}$ (thin dashed line)
at $\tilde{q}=1$.}
\label{fig1} 
\end{figure}

Here we consider the expansion functions of a \emph{homogeneous} Fermi
gas in the unitarity limit and emphasize their universal aspect, which is not known so far. 
As the scattering length diverges, all microscopic scales of the interaction are absent
\cite{houniversality}. For this few-body problem, the only energy
scale is $k_{B}T$ and length scale is the thermal wavelength $\lambda\equiv h/(2\pi mk_{B}T)^{1/2}$.
Dimensional analysis leads to, \begin{equation}
\Delta S_{\sigma\sigma^{\prime},n}({\bf q},\omega,T)=\frac{V}{k_{B}T\lambda^{3}}\Delta\tilde{S}_{\sigma\sigma^{\prime},n}(\tilde{q},\tilde{\omega}),\end{equation}
 where $V$ is the volume, $\tilde{q}=[\hbar^{2}{\bf q}^{2}/(2mk_{B}T)]^{1/2}$,
$\tilde{\omega}=\hbar\omega/(k_{B}T)$, and $\Delta\tilde{S}_{\sigma\sigma^{\prime},n}$
is a dimensionless expansion function. The temperature $T$ is now
implicit in the variables $\tilde{q}$ and $\tilde{\omega}$. This
universal form implies a simple relation between the trapped and homogeneous
expansion function. In a shallow harmonic trap, $V_{Trap}({\bf r})=m(\omega_{x}^{2}x^{2}+\omega_{y}^{2}y^{2}+\omega_{z}^{2}z^{2})/2$,
where $\omega_{0}\equiv(\omega_{x}\omega_{y}\omega_{z})^{1/3}\rightarrow0$,
the system may be viewed as a collection of many cells with a local
chemical potential $\mu({\bf r})=\mu-V_{Trap}({\bf r})$ and fugacity
$z(r)=z\exp[-V_{Trap}({\bf r})/k_{B}T]$, so that the trapped DSF
is given by $\Delta S_{\sigma\sigma^{\prime}}^{(Trap)}\left({\bf q},\omega,T\right)=\int d{\bf r[}\Delta S_{\sigma\sigma^{\prime}}\left({\bf q},\omega,T,{\bf r}\right)/V]$.
Owing to the universal $\tilde{q}$- and $\tilde{\omega}$-dependence
in the expansion functions, the spatial integration can be easily
performed, giving rise to \begin{equation}
\Delta\tilde{S}_{\sigma\sigma^{\prime},n}(\tilde{q},\tilde{\omega})=n^{3/2}\frac{\left(\hbar\omega_{0}\right)^{3}}{\left(k_{B}T\right)^{2}}\Delta S_{\sigma\sigma^{\prime},n}^{(Trap)}({\bf q},\omega,T).\label{UniversalRelation}\end{equation}
 The (non-universal) correction to the above local density approximation
is at the order of $O[(\hbar\omega_{0})^{2}/(k_{B}T)^{2}]$. Eq. (\ref{UniversalRelation})
is vitally important because the calculation of expansion functions
in harmonic traps is much easier than in free space. 

Fig. 1 reports the homogeneous expansion function $\Delta\tilde{S}_{2}=2[\Delta\tilde{S}_{\uparrow\uparrow,2}+\Delta\tilde{S}_{\uparrow\downarrow,2}]$
at three different momenta, using $\Delta S_{\sigma\sigma^{\prime},2}^{(Trap)}$
in Ref. \cite{vedsf} as the input. One observes a quasielastic peak
at $\tilde{\omega}=\tilde{q}^{2}/2$ or $\omega=\hbar{\bf q}^{2}/(4m)$,
as a result of the formation of fermionic pairs. We note that the
third expansion function $\Delta\tilde{S}_{\sigma\sigma^{\prime},3}$
or $\Delta S_{\sigma\sigma^{\prime},3}^{(Trap)}$ can also be calculated
straightforwardly using exact three-fermion solutions \cite{werner2007}.

We may derive sum rules that constrain the universal expansion functions,
using the well-known \textit{f}-sum rules satisfied by DSF: $\int\nolimits _{-\infty}^{+\infty}\omega S_{\uparrow\uparrow}({\bf q},\omega,T)d\omega=N{\bf q}^{2}/(4m)$
\cite{combescot} and $\int\nolimits _{-\infty}^{+\infty}\omega S_{\uparrow\downarrow}({\bf q},\omega,T)d\omega=0$
\cite{guo}. The latter immediately leads to \begin{equation}
\int\nolimits _{-\infty}^{+\infty}\tilde{\omega}\Delta\tilde{S}_{\uparrow\downarrow,n}(\tilde{q},\tilde{\omega})d\tilde{\omega}=0.\end{equation}
 On the other hand, virial expansion of the total number of fermions
$N$ implies that \begin{equation}
\int\nolimits _{-\infty}^{+\infty}\tilde{\omega}\Delta\tilde{S}_{\uparrow\uparrow,n}(\tilde{q},\tilde{\omega})d\tilde{\omega}=n\tilde{q}^{2}\Delta b_{n},\end{equation}
 where $\Delta b_{n}$ is the \textit{n}-th virial coefficient and
in the unitarity limit $\Delta b_{2}=1/\sqrt{2}$ and $\Delta b_{3}\simeq-0.355$
\cite{lhdprl2009,ensEoS}. 

At large momentum, the spin-antiparallel static structure factor satisfies
the Tan relation \cite{sutTanRelation}, $\int S_{\uparrow\downarrow}({\bf q},\omega,T)d\omega\simeq{\cal I}/(8\hbar q)$.
This indicates a virial expansion of the contact: ${\cal I}=16\pi^{2}V\left[z^{2}c_{2}+z^{3}c_{3}+\cdots\right]/\lambda^{4}$,
where the contact coefficients $c_{n}$ are given by, \begin{equation}
\Delta\tilde{S}_{\uparrow\downarrow,n}(\tilde{q}\gg1)\equiv\int\nolimits _{-\infty}^{+\infty}\Delta\tilde{S}_{\uparrow\downarrow,n}(\tilde{q},\tilde{\omega})d\tilde{\omega}=\frac{\pi^{3/2}c_{n}}{\tilde{q}}.\label{contactSumRule}\end{equation}
 The expansion of the contact was alternatively obtained using Tan's
adiabatic sweep relation \cite{vecontact}. In the unitarity limit,
it was shown that $c_{2}=1/\pi$ and $c_{3}\simeq-0.141$ \cite{vecontact}.
In the same limit of large momentum, the spin-parallel static structure
factor is nearly unity so that $\int S_{\uparrow\uparrow}({\bf q},\omega,T)d\omega\simeq N/(2\hbar)$
\cite{vedsf,sutTanRelation}. This leads to
\begin{equation}
\Delta\tilde{S}_{\uparrow\uparrow,n}(\tilde{q}\gg1)\equiv\int\nolimits _{-\infty}^{+\infty}\Delta\tilde{S}_{\uparrow\uparrow,n}(\tilde{q},\tilde{\omega})d\tilde{\omega}=n\Delta b_{n}.
\end{equation}
For the second expansion function, $\Delta\tilde{S}_{\sigma\sigma^{\prime},2}$, we have checked numerically that all the 
above mentioned sum rules are strictly satisfied. 

\section{Wilson coefficient of DSF}

We now turn to the large momentum/frequency limits, where the OPE Eq. (1) is assumed to be applicable.
It is clear from the equation that the Wilson coefficient determines
entirely the DSF at large $({\bf q},\omega)$ as far as the many-body
contact is known.

\begin{figure}
\onefigure[clip,width=0.48\textwidth]{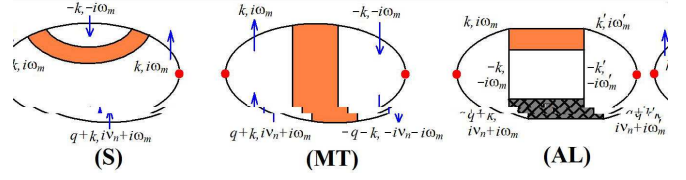} 
\caption{(Color on-line) Diagrammatic contributions to the interaction dynamic
susceptibility. The self-energy (S) and Maki-Thompson (MT) diagrams
contribute to $\Delta\chi_{\uparrow\uparrow}({\bf r},\tau)$ and $\Delta\chi_{\uparrow\downarrow}({\bf r},\tau)$,
respectively, while the Aslamazo-Larkin diagram (AL) contributes to
both. The shadow in the diagrams represents the contact $\mathcal{I}$.
The crossed part in the diagram (AL) is the vertex.}
\label{fig2} 
\end{figure}

At \textit{T}=0, the Wilson coefficient $W_{\sigma\sigma^{\prime}}$
can be calculated using Feynman diagrams \cite{son} for dynamic susceptibility
$\chi_{\sigma\sigma^{\prime}}({\bf r},\tau)=-\left\langle T_{\tau}\hat{\rho}_{\sigma}({\bf r},\tau)\hat{\rho}_{\sigma^{\prime}}({\bf 0},0)\right\rangle $,
as $S_{\sigma\sigma^{\prime}}({\bf q},\omega)=-\mathop{\rm Im}\chi_{\sigma\sigma^{\prime}}({\bf q},\omega)/[\pi(1-e^{-\hbar\omega/k_{B}T})]$.
In the limit of $({\bf q},\omega)\rightarrow\infty$, the contributing
diagrams to $\chi_{\sigma\sigma^{\prime}}({\bf q},\omega)$ are sketched
in Fig. 2 \cite{son}. Diagrammatically, the contact may be identified
as the vertex function at short distance and time \cite{vecontact,haussmann}:
${\cal I}=-m^{2}\Gamma({\bf r}={\bf 0},\tau=0^{-})/\hbar^{4}$. Therefore,
in the diagrams the shadow part of the vertex function $\Gamma({\bf r}={\bf 0},\tau=0^{-})$
represents the contact ${\cal I}$. These diagrams are well-known
in condensed matter community. In the context of calculating the change
in conductivity due to conducting fluctuations, the last two diagrams
are called the Maki-Thompson (MT) \cite{MT} and Aslamazov-Larkin
(AL) contributions \cite{AL} respectively, while first diagram gives
the self-energy (S) correction. At zero temperature, we calculate
these diagrams in vacuum at $\mu=0$ and obtain that $W_{\uparrow\uparrow}^{T=0}=(f_{S}-f_{AL})/(4\pi^{2}\sqrt{m\hbar}\omega^{3/2})$
and $W_{\uparrow\downarrow}^{T=0}=(f_{MT}-f_{AL})/(4\pi^{2}\sqrt{m\hbar}\omega^{3/2})$,
where, 
\begin{eqnarray}
f_{S}& = &\frac{\sqrt{1-x/2}}{\left(1-x\right)^{2}}, \nonumber\\
f_{MT}& = &\frac{1}{\sqrt{2x}}\ln\frac{1+\sqrt{2x-x^{2}}}{\left|1-x\right|}, \nonumber \\
f_{AL}& = &\frac{1}{2x\sqrt{1-x/2}}\left[\ln^{2}\frac{1+\sqrt{2x-x^{2}}}{\left|1-x\right|}-\pi^{2}\Theta(x-1)\right], \nonumber
\end{eqnarray}
 $x\equiv\hbar^{2}{\bf q}^{2}/(2m\hbar\omega)$, and $\Theta$ is
the step function. These results agree with the previous calculations
by Son and Thompson \cite{son}, although there the spin parallel
and antiparallel DSFs were not treated separately. At small ${\bf q}^{2}/\omega$,
we find that the spin parallel and antiparallel DSFs have the tail
\begin{equation}
W_{\uparrow\uparrow}^{T=0}=-W_{\uparrow\downarrow}^{T=0}=\frac{\hbar^{1/2}{\bf q}^{2}}{12\pi^{2}m^{3/2}\omega^{5/2}}.\label{WilsonTail}
\end{equation}
 This prediction shows that for $\omega\rightarrow\infty$ the spin
dependent DSFs decay an order slower in magnitude than the total DSF.
The latter is proportional to $q^{4}/\omega^{7/2}$, as shown in Refs.
\cite{son} and \cite{taylor}. The faster decay in the total dynamic structure factor is 
due to the cancellation of the leading terms in $W_{\uparrow\uparrow}^{T=0}$ and $W_{\uparrow\downarrow}^{T=0}$.

\begin{figure}
\onefigure[clip,width=0.40\textwidth]{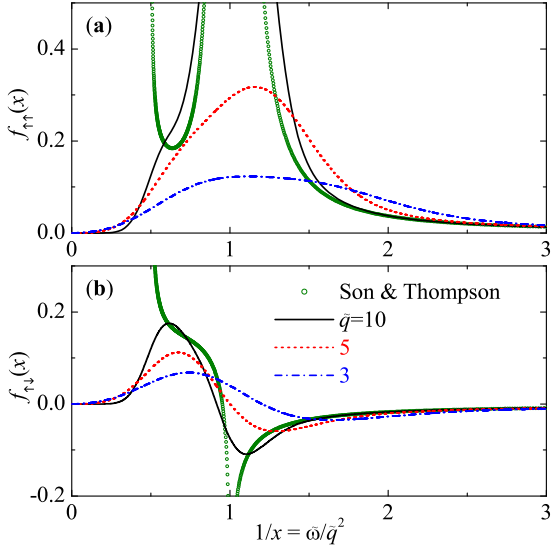} 
\caption{(Color on-line) $f_{\sigma\sigma^{\prime}}=\sqrt{m\hbar}\omega^{3/2}(z^{2}/{\cal I}_{2})\Delta S_{\sigma\sigma^{\prime},2}$
at $\tilde{q}=3$, $5$, and $10$. With increasing momentum and/or
frequency, $f_{\sigma\sigma^{\prime}}$ approaches smoothly to the
$T=0$ result by Son and Thompson \cite{son}.}
\label{fig3} 
\end{figure}

It is not clear how to obtain the finite temperature Wilson coefficient
using diagrammatic technique, since the finite temperature diagrams
involve many-body process in medium. However, Eq. (1) gives the hint.
It has a strong constraint on the expansion functions of DSF. As $W_{\sigma\sigma^{\prime}}$
involves only the few-body physics and hence does not contain the
fugacity $z$, a count of the term $z^{n}$ on both sides of Eq. (1)
leads to 
\begin{eqnarray}
\Delta S_{\sigma\sigma^{\prime},n}\left({\bf q},\omega,T\right) & = & \left(c_{n}/c_{2}\right)\Delta S_{\sigma\sigma^{\prime},2}\left({\bf q},\omega,T\right),\label{HigherOrderExpansion}\\
W_{\sigma\sigma^{\prime}}\left({\bf q},\omega,T\right) & = & \left(z^{2}/{\cal I}_{2}\right)\Delta S_{\sigma\sigma^{\prime},2}\left({\bf q},\omega,T\right),\label{WilsonCoefficient}
\end{eqnarray}
where ${\cal I}_{2}=z^{2}16\pi^{2}Vc_{2}/\lambda^{4}$ is the contact
up to the second order expansion \cite{vecontact}. Therefore, the
Wilson coefficient is given by the second expansion function. This
result is obtained by applying the OPE and virial expansion method.
As a result, in principle it should be valid at temperatures above
the superfluid transition. However, we may expect that it holds at
all temperatures, as both the Wilson coefficient and second expansion function are irrelevant to the many-body pairing 
in the superfluid phase. The many-body effect enters through the many-body 
parameter of contact only. 

Eq. (\ref{WilsonCoefficient}) is the main result of this Letter.
At a first glance, it may be a suprising result. However, it could
be understood from the proportionality shown in Eq. (\ref{HigherOrderExpansion}).
On the other hand, a rigorous proof of Eq. (\ref{HigherOrderExpansion})
at large $q$ and $\omega$ justifies the use of the OPE method. We
note that, for the two-component Fermi gas, our result shows that
the Wilson coefficient is determined by the two-body physics solely.
This is in agreement with previous work. For example, in Ref. \cite{bratten2010}
a two-body scattering amplitude was used to calculate the zero temperature
Wilson coefficient for the rf-spectroscopy of a strongly interacting
Fermi gas. However, in general cases where three or four-body physics
come into play, we anticipate that the Wilson coefficient should be
related to the higher order virial expansion function. In that case,
new universal relations with new many-body {}``contact'' parameters
would appear. 

In Fig. 3 we check the validity of Eq. (\ref{WilsonCoefficient})
at $T=0$, by calculating $\sqrt{m\hbar}\omega^{3/2}(z^{2}/{\cal I}_{2})\Delta S_{\sigma\sigma^{\prime},2}$
at different momenta. With decreasing temperature $T$
or increasing $\tilde{q}\propto q/\sqrt{T}$, it approaches gradually
to $\sqrt{m\hbar}\omega^{3/2}W_{\sigma\sigma^{\prime}}^{T=0}$. This
confirms numerically that Eq. (\ref{WilsonCoefficient}) holds at 
zero temperature. Moreover, at high temperatures where the fugacity is small,
to a good approximation we have the interaction DSF $\Delta S_{\sigma\sigma^{\prime}} \simeq z^2 \Delta S_{\sigma\sigma^{\prime},2}$. 
As the contact ${\cal I} \simeq {\cal I}_{2}$ at high $T$, it is trivial to confirm Eq. (\ref{WilsonCoefficient}). 
Note that, in the limit of large frequency, the tail $\omega^{-5/2}$ of $W_{\uparrow\uparrow}^{T=0}$ 
and $W_{\uparrow\downarrow}^{T=0}$ is fairly evident in the second order virial expansion functions.

\section{Universal DSF at large $({\bf q},\omega)$}

In this limit, using Eqs. (\ref{OPE}) and (\ref{WilsonCoefficient}) the DSF is approximated by 
\begin{equation}
S_{\sigma\sigma^{\prime}}\simeq S_{\sigma\sigma^{\prime}}^{(1)}\left({\bf q},\omega,T\right)+ \frac{ {\cal I}\lambda^{4}}{16\pi V}\Delta S_{\sigma\sigma^{\prime},2}\left({\bf q},\omega,T\right).\label{LargeQwDSF}
\end{equation}
This approximate DSF should be \emph{quantitatively} accurate for sufficiently large momentum
and frequency. It holds at all temperatures and the many-body effect is respected by
the contact. However, the momentum $q$ should be larger than a critical momentum $q_c \gg max(\lambda^{-1}, k_F)$ in order to 
validate the use of the OPE equation (\ref{OPE}). Here, $k_{F}$ is the Fermi wavevector. 
Quantitatively, an estimate of $q_c$ requires the calculation of $\Delta S_{\sigma\sigma^{\prime},n>2}$ and the 
examination of Eq. (\ref{HigherOrderExpansion}).  We note that Eq. (\ref{LargeQwDSF}) may not be reliable at small frequency, $\omega\sim0$. 
As a result, the structure factor sum-rules may not be strictly satisfied. We note also that, in the limit of high temperatures
where the contact ${\cal I} \simeq {\cal I}_{2}$, the prefactor of the $\Delta S_{\sigma\sigma^{\prime},2}$ term is 
$z^2$. Therefore, the approximate DSF reduces back to the virial expansion up to the second order. 
In this high-$T$ limit, Eq. (\ref{LargeQwDSF}) is valid for arbitrary $q$ and $\omega$.

\begin{figure}
\onefigure[clip,width=0.48\textwidth]{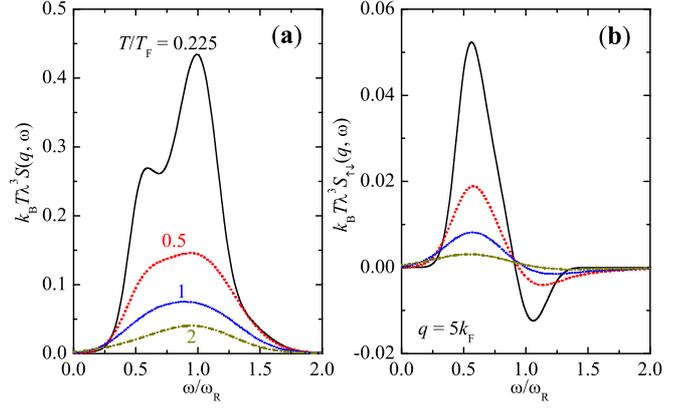} 
\caption{(Color on-line) Universal total DSF (a) and spin-antiparallel (b) DSF
at $T/T_{F}=0.225$, $0.5$, $1$, and $2$, calculated using Eq.
(\ref{LargeQwDSF}). We use a pair fluctuation theory to determine
the fugacity $z$ and contact $\mathcal{I}$ \cite{vecontact}.}
\label{fig4} 
\end{figure}

We present in Fig. 4 the temperature 
dependence of the approximation DSF of a normal, homogeneous unitary Fermi gas at $q=5k_{F}$, by assuming 
that $q_c \sim 5k_{F}$. The many-body contact and fugacity are calculated by using a strong-coupling pair fluctuation
theory \cite{vecontact}, which is shown to be accurate for describing the unitary equation of state. 
Close to the superfluid transition temperature, a quasielastic peak clearly emerges at the recoil energy for pairs, 
$\omega=\hbar{\bf q}^{2}/(4m)$, in agreement with the low-temperature experimental observation \cite{sutbragg}.

\section{Observation of universal DSF at large frequency}

Eq. (\ref{LargeQwDSF}) indicates
that at large\textit{\ }$({\bf q},\omega)$, the interaction DSF
of a unitary Fermi gas depends on the reduced moment $\tilde{q}=q\lambda/\sqrt{4\pi}$
and reduced frequency $\tilde{\omega}$ only. This universal dependence
could be examined using large-momentum Bragg spectroscopy \cite{sutbragg,sutTanRelation},
with varying momentum and temperature while keeping $\tilde{q}$ invariant.
One can also extract experimentally the universal second expansion
function $\Delta\tilde{S}_{\sigma\sigma^{\prime},2}(\tilde{q},\tilde{\omega})$
since the contact ${\cal I}$ can be determined independently using
the \textit{f}-sum rule \cite{sutTanRelation}. These predictions
break down below the critical momentum $q_{c}$. Note that, by tuning the transferred
momentum in Bragg beams, the value of $q_{c} \gg max(\lambda^{-1}, k_F) $ might be determined
experimentally.

\section{Conclusion}

We have studied the finite temperature dynamic
structure factor of a homogeneous unitary Fermi gas, using quantum
virial expansion and operator product expansion. The universal second
order expansion function has been calculated and related to the Wilson
coefficient at large momentum $q$. As a result, in that limit the
thermal wavelength $\lambda$ becomes the only length scale and the
interaction dynamic structure factor should depend universally on
a reduced momentum $q\lambda/\sqrt{4\pi}$. We have proposed that
Bragg spectroscopy with large transferred momentum should be able
to confirm this universal dependence. Our results can be extended
to other dynamical properties of a strongly correlated Fermi gas,
such as the rf-spectrum and single-particle spectral function. 

It is an interesting chanllenge to derive from three- and four-fermion
solutions \cite{werner2007,daily} new universal relations involving
a many-body parameter like Tan's contact. The determination of Wilson
coefficients in that case should be difficult. Our method of calculating
higher-order virial expansion function would give the most natural
and convenient way to obtain the Wilson coefficient.

\acknowledgments
We thank P. D. Drummond, P. Hannaford, E. D. Kuhnle, S. Tan, and C.
J. Vale for fruitful discussions. This research was supported by the
ARC Centre of Excellence, ARC Discovery Project No. DP0984522 and
No. DP0984637.

\end{document}